\documentclass[epj]{svjour}

\usepackage[centertags]{amsmath}
\allowdisplaybreaks[1]

\usepackage{amsbsy}
\usepackage{amsfonts}
\usepackage{amssymb}
\usepackage[dvips]{graphicx}
\usepackage{url}

\begin{document}

\title{Status of Neutrino Masses and Mixing}

\author{Carlo Giunti}

\institute{INFN, Sezione di Torino, and Dipartimento di Fisica Teorica,
Universit\`a di Torino,
Via P. Giuria 1, I--10125 Torino, Italy}

\date{Talk presennted at
HEP2003, 17--23 July 2003, Aachen, Germany}

\abstract{The experimental evidences in favor
of oscillations of solar (and KamLAND)
and atmospheric (and K2K)
neutrinos are briefly reviewed
and accommodated in the framework of three-neutrino mixing.
The implications for the values of neutrino masses are discussed
and
the bounds on the absolute scale of neutrino masses from Tritium $\beta$-decay
and cosmological data are reviewed.
Finally, we discuss
the implications of three-neutrino mixing for neutrinoless
double-$\beta$ decay.
}

\maketitle



\begin{figure}
\begin{center}
\begin{minipage}[t]{0.50\textwidth}
\begin{center}
\includegraphics*[bb=22 81 513 676, width=0.60\textwidth]{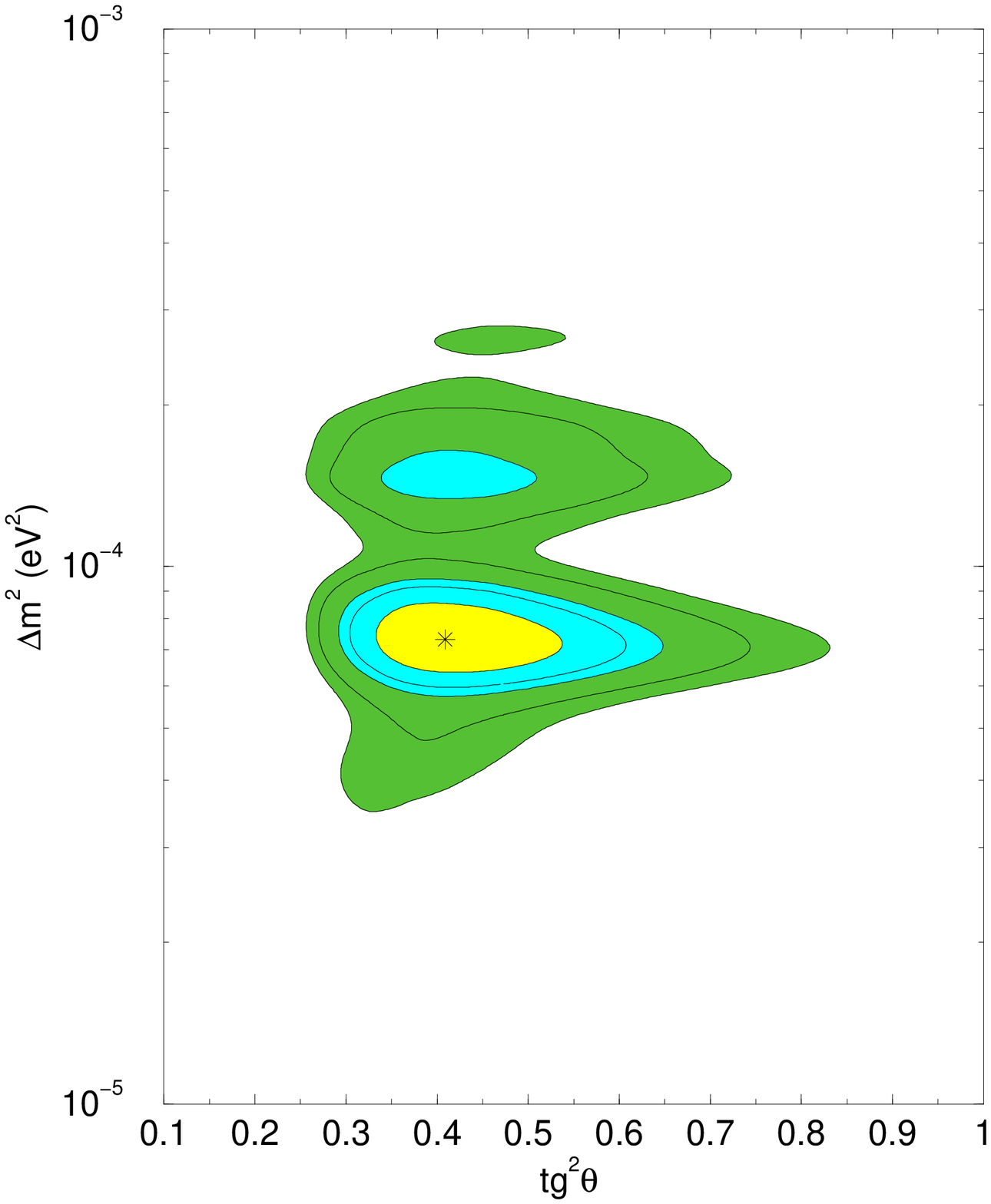}
\end{center}
\end{minipage}
\end{center}
\caption{ \label{deholanda-0212270-f04}
Allowed
68.3\%,
90\%,
95\%,
99\%,
99.73\%
C.L.
regions
obtained from the global analysis of
solar and KamLAND data.
The best-fit point is marked by a star.
Figure from Ref.~\protect\cite{hep-ph/0212270}.
}
\end{figure}

About one year ago,
the observation of solar neutrinos
through neutral-current and charged-current
reactions
allowed the SNO experiment \cite{Ahmad:2002jz}
to solve the long-standing solar neutrino problem
in favor of the existence of $\nu_e \to \nu_\mu, \nu_\tau$
transitions.
The global analysis of all solar neutrino data
in terms of the most natural hypothesis of neutrino oscillations
favored the so-called Large Mixing Angle (LMA)
region
with
a squared-mass difference
$
2 \times 10^{-5}
\lesssim
\Delta{m}^2_{\mathrm{SUN}}
\lesssim
4 \times 10^{-4}
$
(we measure squared-mass differences in units of eV$^2$)
and a large effective mixing angle
$ 0.2 \lesssim \tan^2\vartheta \lesssim 0.9 $
(see Ref.~\cite{hep-ph/0301276}).
A spectacular proof of the correctness of the LMA region
has been obtained at the end of last year in
the KamLAND long-baseline $\bar\nu_e$ disappearance experiment
\cite{hep-ex/0212021},
in which a suppression of
$ 0.611 \pm 0.085 \pm 0.041 $
of the $\bar\nu_e$ flux
produced by nuclear reactors at an average distance of about 180 km
was observed.
The allowed regions of the effective neutrino oscillation parameters
obtained from the global analysis of solar and KamLAND neutrino data
are shown in Fig.~\ref{deholanda-0212270-f04}
\cite{hep-ph/0212270}.
The effective squared-mass difference
$\Delta{m}^2_{\mathrm{SUN}}$
is constrained in one of the two ranges
\cite{hep-ph/0212129}
\begin{subequations}
\label{001}
\begin{align}
\text{LMA-I:}
\quad
\null & \null
5.1 \times 10^{-5}
<
\Delta{m}^2_{\mathrm{SUN}}
<
9.7 \times 10^{-5}
\,,
\label{001a}
\\
\text{LMA-II:}
\quad
\null & \null
1.2 \times 10^{-4}
<
\Delta{m}^2_{\mathrm{SUN}}
<
1.9 \times 10^{-4}
\,,
\label{001b}
\end{align}
\end{subequations}
at 99.73\% C.L.,
with best-fit value
$ \Delta{m}^{2\,\mathrm{bf}}_{\mathrm{SUN}} \simeq 6.9 \times 10^{-5} $
in the LMA-I region
(see also Ref.~\cite{hep-ph/0301276} and references therein).
The effective solar mixing angle $\vartheta_{\mathrm{SUN}}$
is constrained at 99.73\% C.L. in the interval
\cite{hep-ph/0212129}
\begin{equation}
0.29 < \tan^2 \vartheta_{\mathrm{SUN}} < 0.86
\,,
\label{002}
\end{equation}
with best-fit value
$ \tan^2 \vartheta_{\mathrm{SUN}}^{\mathrm{bf}} \simeq 0.46 $.

\begin{figure}
\begin{center}
\begin{minipage}[t]{0.50\textwidth}
\begin{center}
\includegraphics*[bb=18 254 278 550, width=0.60\textwidth]{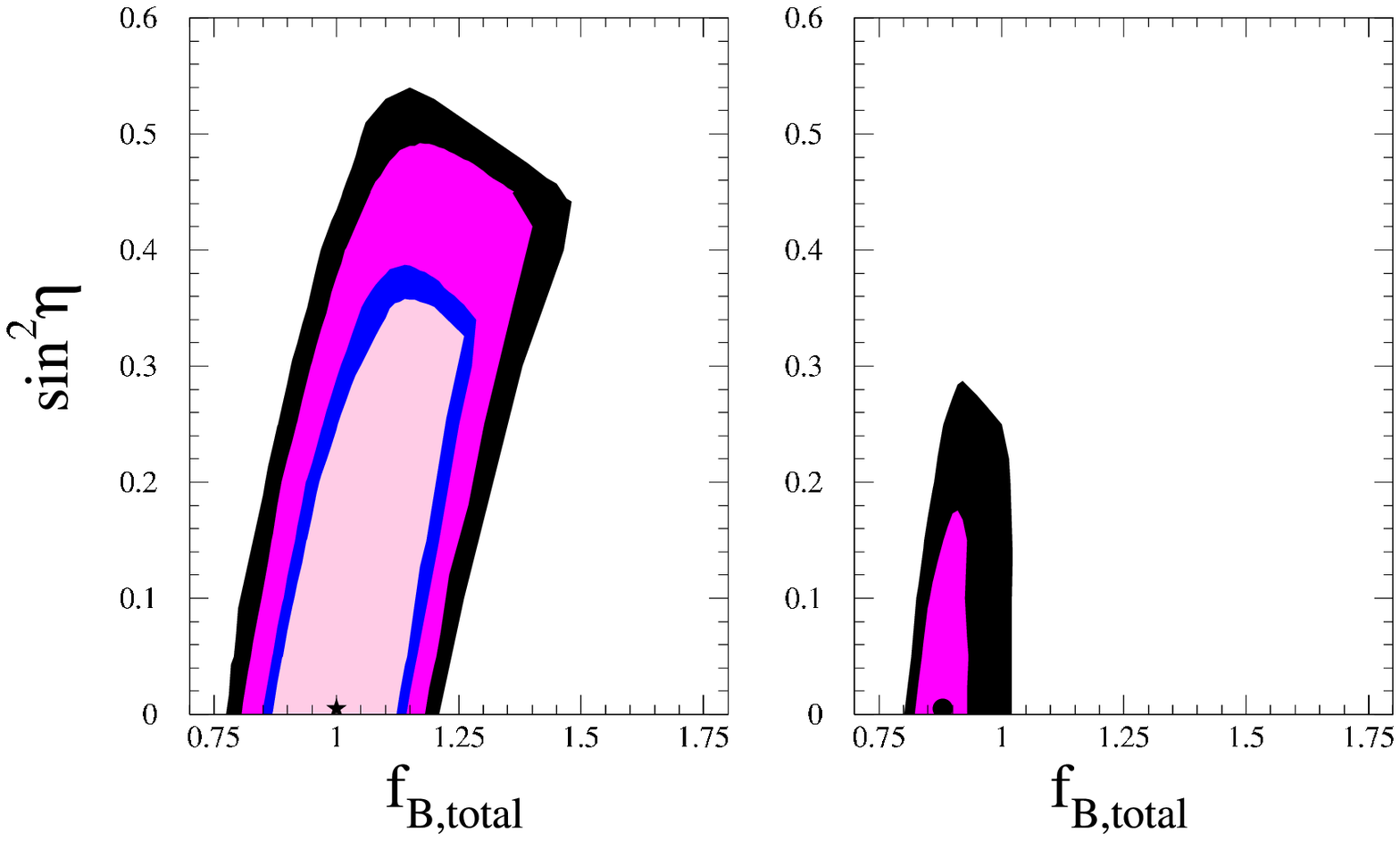}
\end{center}
\end{minipage}
\end{center}
\caption{ \label{bahcall-0212147-f06}
Allowed
90\%, 95\%, 99\%, 99.73\% C.L.
regions obtained from the global analysis of
solar and KamLAND data.
The best-fit point is marked by a star.
Figure from Ref.~\protect\cite{hep-ph/0212147}.
}
\end{figure}

Transitions of solar $\nu_e$'s into sterile states are disfavored by
the data.
Figure~\ref{bahcall-0212147-f06}
\cite{hep-ph/0212147}
shows the allowed regions in the
$\mathrm{f_{B,total}}$--$\sin^2\eta$ plane,
where
$\mathrm{f_{B,total}}=\Phi_{^8\mathrm{B}}/\Phi_{^8\mathrm{B}}^{\mathrm{SSM}}$
is the ratio of the $^8\mathrm{B}$ solar neutrino flux
and its value predicted by the Standard Solar Model (SSM)
\cite{Bahcall:2000nu}.
The parameter $\sin^2\eta$ quantifies the fraction of
solar $\nu_e$'s that transform into sterile $\nu_s$:
$\nu_e \to \cos\eta \, \nu_a + \sin\eta \, \nu_s $,
where $\nu_a$ are active neutrinos.
From Fig.~\ref{bahcall-0212147-f06}
it is clear that there is a correlation between
$\mathrm{f_{B,total}}$
and
$\sin^2\eta$,
which is due to the constraint on the total flux of
$^8\mathrm{B}$ active neutrinos given by the SNO neutral-current measurement:
disappearance into sterile states is possible only if the
$^8\mathrm{B}$ solar neutrino flux
is larger than the SSM prediction.
The allowed ranges for
$\Phi_{^8\mathrm{B}}$
and
$\sin^2\eta$
are
\cite{hep-ph/0212147}
\begin{equation}
\Phi_{^8\mathrm{B}} = 1.00 \pm 0.06 \, \Phi_{^8\mathrm{B}}^{\mathrm{SSM}}
\,,
\quad
\sin^2\eta < 0.52
\,.
\label{014}
\end{equation}
The allowed interval for
$\Phi_{^8\mathrm{B}}$
shows a remarkable agreement of the
data with the SSM,
independently from possible $\nu_e\to\nu_s$ transitions.

In the future it is expected that
the KamLAND experiment will allow to distinguish between
the LMA-I and LMA-II regions,
reaching a relatively high accuracy in the determination of
$\Delta{m}^2_{\mathrm{SUN}}$
\cite{Inoue:2003qs},
whereas new low-energy solar neutrino experiments
or a new dedicated reactor neutrino experiment
are needed in order to improve significantly our knowledge
of the solar effective mixing angle $\vartheta_{\mathrm{SUN}}$
\cite{hep-ph/0302243,Bahcall:2003ce,hep-ph/0306017}.

\begin{figure}
\begin{center}
\begin{minipage}[t]{0.50\textwidth}
\begin{center}
\includegraphics*[bb=42 299 86 509, scale=0.75]{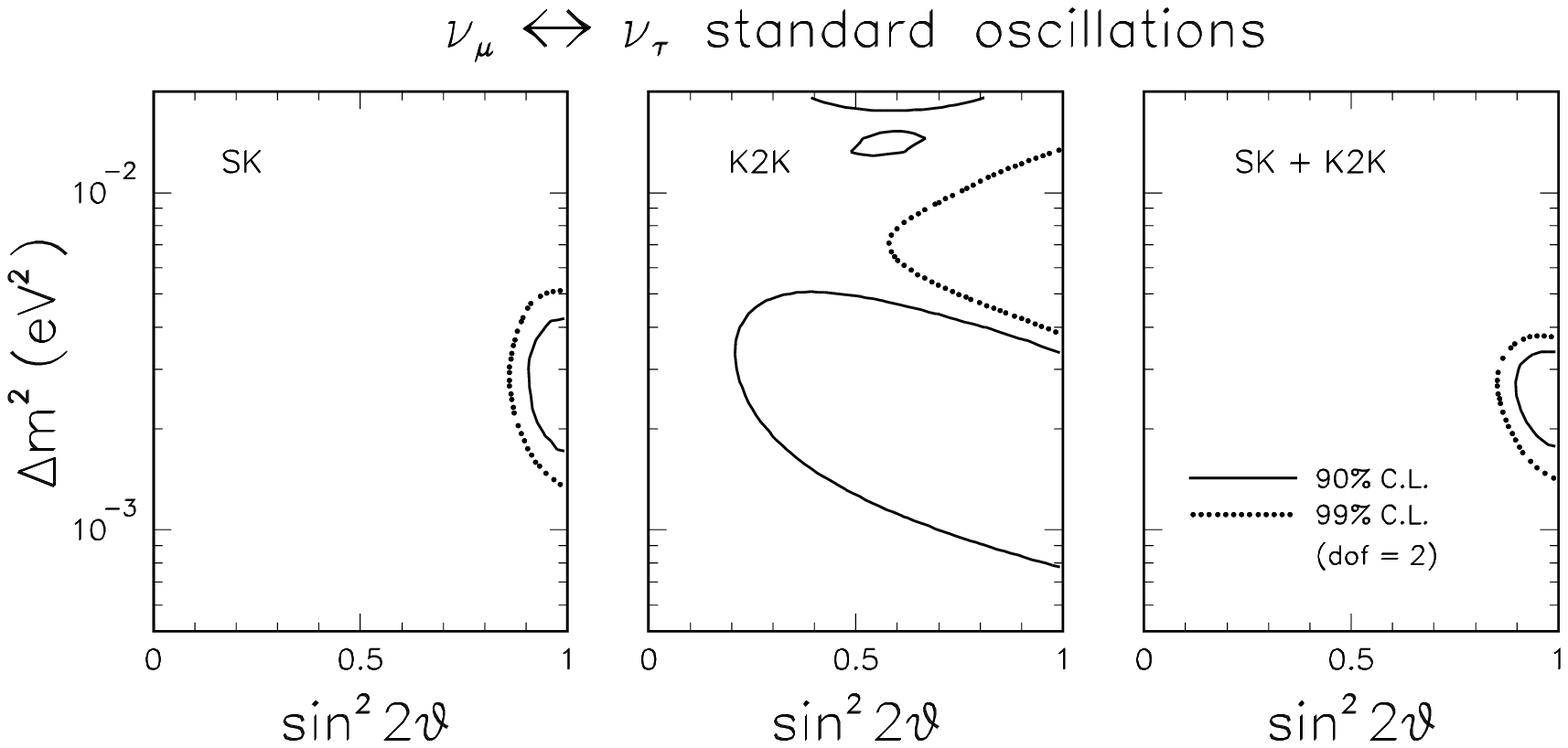}
\includegraphics*[bb=398 299 536 509, scale=0.75]{fig/fogli-0303064-f01.eps.pdf.eps}
\end{center}
\end{minipage}
\end{center}
\caption{ \label{fogli-0303064-f01}
Allowed region
obtained from the analysis of
Super-Kamiokande atmospheric and K2K data
in terms of
$\nu_\mu\to\nu_\tau$
oscillations.
Figure from Ref.~\protect\cite{hep-ph/0303064}.
}
\end{figure}

In 1998 the Super-Kamiokande Collaboration
\cite{Fukuda:1998mi}
discovered
the up-down asymmetry of high-energy events generated
by atmospheric $\nu_\mu$'s,
providing a model independent proof
of atmospheric $\nu_\mu$ disappearance.
At the end of 2002
the long-baseline K2K experiment
\cite{Ahn:2002up}
confirmed the neutrino oscillation
interpretation of the atmospheric neutrino anomaly
observing the disappearance of accelerator $\nu_\mu$'s
at a distance of 250 km from the source.
The data of atmospheric and K2K experiments
are well fitted by
$\nu_\mu \to \nu_\tau$
transitions generated by
the squared-mass difference
$\Delta{m}^2_{\mathrm{ATM}}$
in the 99.73\% C.L. range
\cite{hep-ph/0303064}
\begin{equation}
1.4 \times 10^{-3}
<
\Delta{m}^2_{\mathrm{ATM}}
<
5.1 \times 10^{-3}
\,,
\label{003}
\end{equation}
with best-fit value
$
\Delta{m}^{2\,\mathrm{bf}}_{\mathrm{ATM}}
\simeq
2.6 \times 10^{-3}
$.
The best-fit effective atmospheric mixing
is maximal,
$
\sin^2 2 \vartheta_{\mathrm{ATM}}^{\mathrm{bf}} \simeq 1
$,
with the 99.73\% C.L. lower bound
\cite{hep-ph/0303064}
\begin{equation}
\sin^2 2 \vartheta_{\mathrm{ATM}} > 0.86
\,.
\label{004}
\end{equation}
Figure~\ref{fogli-0303064-f01}
\cite{hep-ph/0303064}
shows the
allowed region
obtained from the analysis of
Super-Kamiokande atmospheric and K2K data.

Transitions of
atmospheric $\nu_\mu$'s into $\nu_e$'s
or sterile states
are disfavored.
The fraction
$\sin^2\xi$ of atmospheric $\nu_\mu$'s
that transform into sterile $\nu_s$
($\nu_\mu \to \cos\xi \, \nu_\tau + \sin\xi \, \nu_s $)
is limited by \cite{Nakaya:2002ki}
\begin{equation}
\sin^2\xi < 0.19
\quad
\text{(90\% C.L.)}
\,.
\label{0041}
\end{equation}

\begin{figure}
\begin{center}
\begin{minipage}[t]{0.50\textwidth}
\setlength{\tabcolsep}{0cm}
\begin{tabular}{lr}
\includegraphics*[bb=181 466 428 775, width=0.49\linewidth]{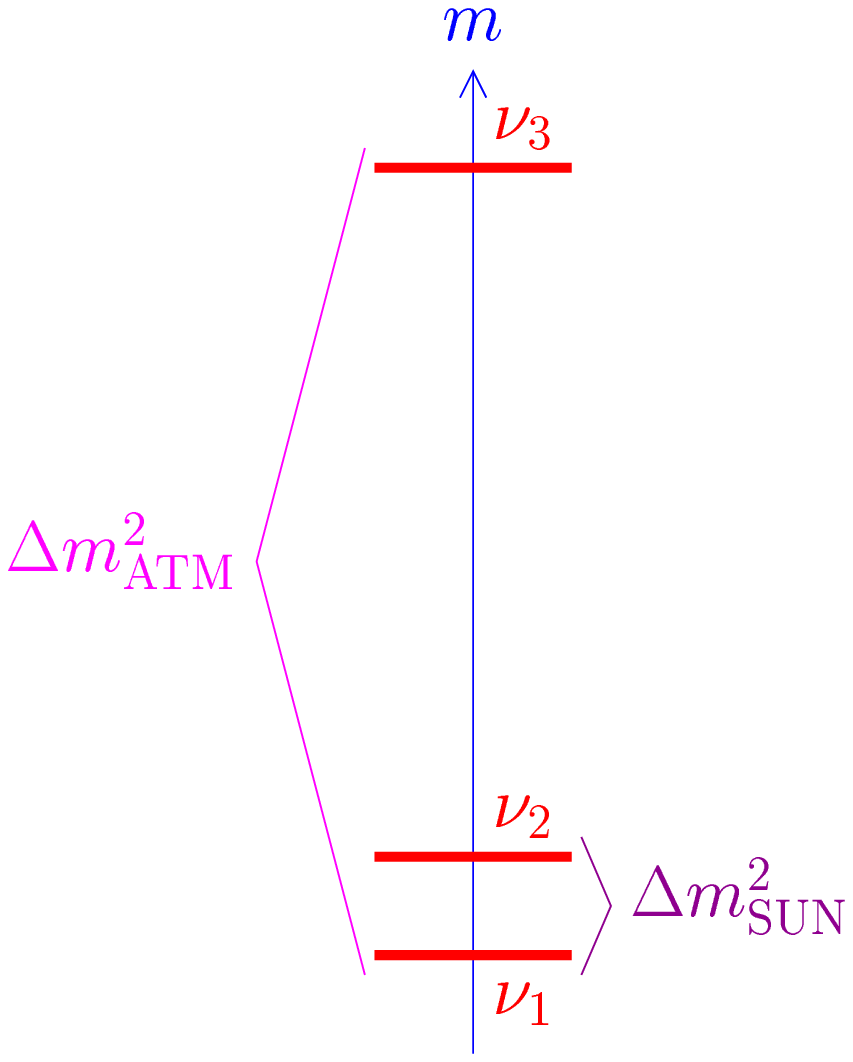}
&
\includegraphics*[bb=183 466 432 775, width=0.49\linewidth]{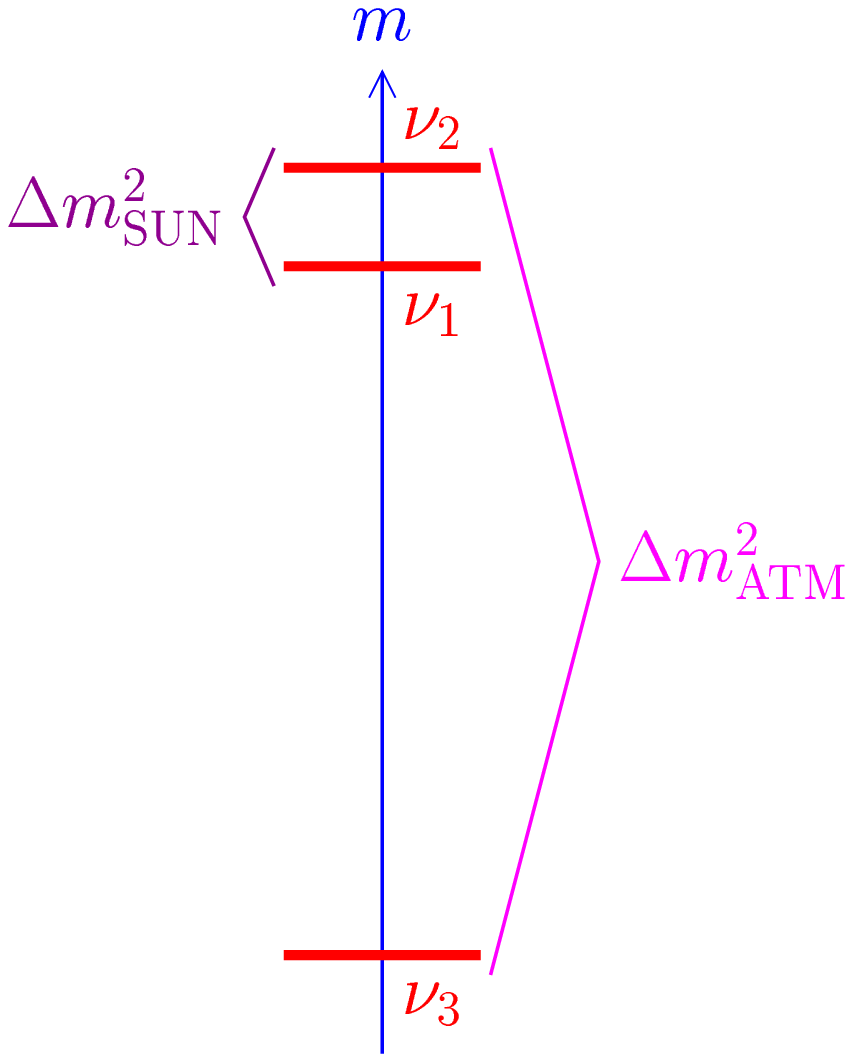}
\\
\textsf{normal}
&
\textsf{inverted}
\end{tabular}
\end{minipage}
\end{center}
\caption{ \label{3nu}
The two three-neutrino schemes allowed by the hierarchy
$\Delta{m}^2_{\mathrm{SUN}} \ll \Delta{m}^2_{\mathrm{ATM}}$.
}
\end{figure}

The solar and atmospheric evidences of neutrino oscillations
are nicely accommodated in the minimal framework of three-neutrino mixing,
in which the three flavor neutrinos
$\nu_e$,
$\nu_\mu$,
$\nu_\tau$
are unitary linear combinations of
three neutrinos
$\nu_1$,
$\nu_2$,
$\nu_3$
with masses
$m_1$,
$m_2$,
$m_3$,
respectively
(see Ref.~\cite{BGG-review-98}).
Figure~\ref{3nu}
shows the two three-neutrino schemes
allowed by the observed hierarchy
of squared-mass differences,
$\Delta{m}^2_{\mathrm{SUN}} \ll \Delta{m}^2_{\mathrm{ATM}}$,
with
the massive neutrinos labeled in order to have
\begin{equation}
\Delta{m}^2_{\mathrm{SUN}}
=
\Delta{m}^2_{21}
\,,
\quad
\Delta{m}^2_{\mathrm{ATM}}
\simeq
|\Delta{m}^2_{31}|
\simeq
|\Delta{m}^2_{32}|
\,.
\label{006}
\end{equation}
The two schemes
in Fig.~\ref{3nu} are usually called
``normal''
and
``inverted'',
because in the normal scheme the smallest
squared-mass difference is generated by the two lightest neutrinos
and a natural neutrino mass hierarchy can be realized if
$m_1 \ll m_2$,
whereas in the inverted scheme the smallest
squared-mass difference is generated
by the two heaviest neutrinos,
which are almost degenerate for any value of the lightest neutrino mass $m_3$.

\begin{figure*}
\begin{minipage}[t]{0.47\textwidth}
\begin{center}
\includegraphics*[bb=118 377 465 702, width=0.99\textwidth]{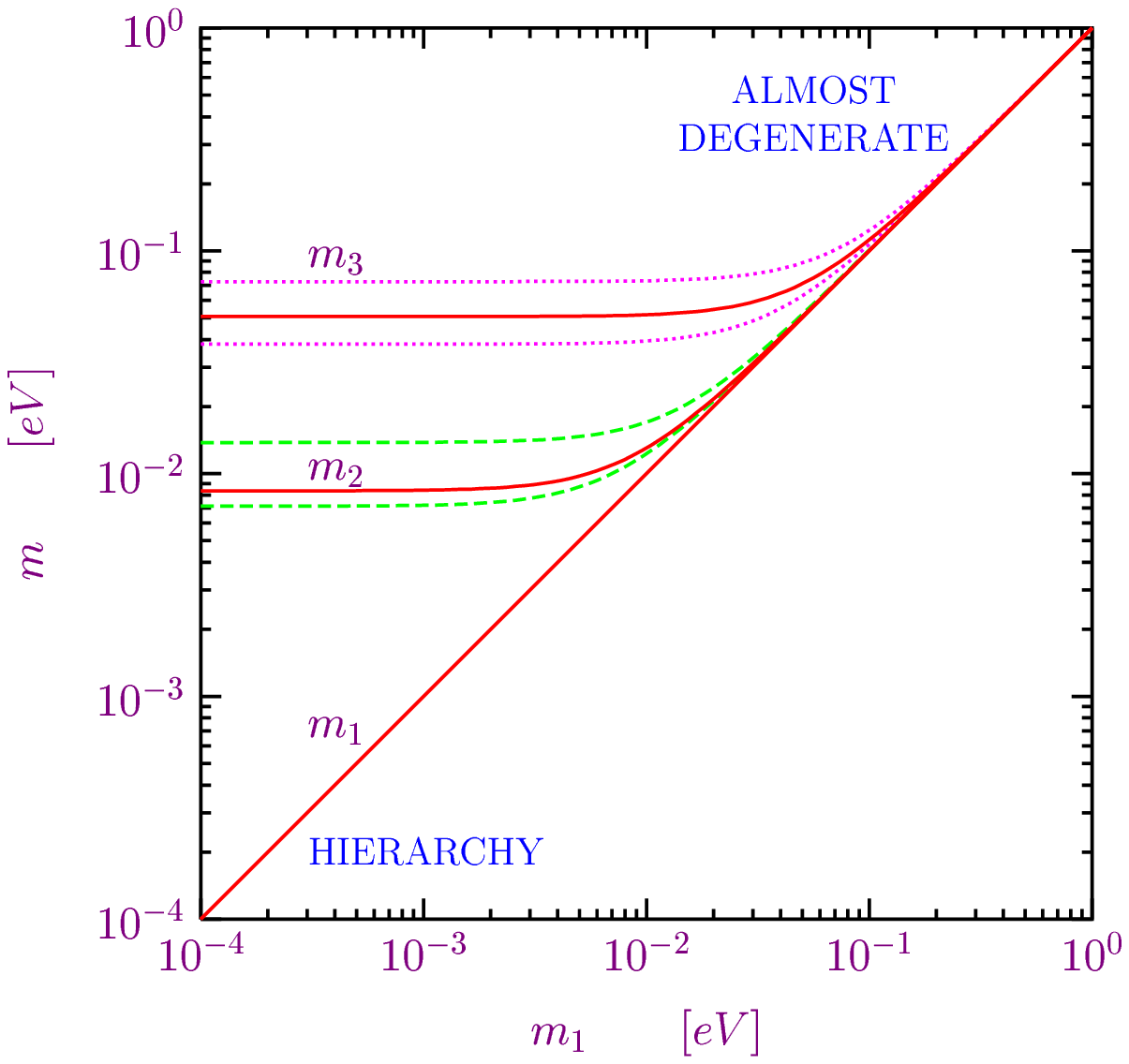}
\end{center}
\end{minipage}
\hfill
\begin{minipage}[t]{0.47\textwidth}
\begin{center}
\includegraphics*[bb=118 427 465 753, width=0.99\textwidth]{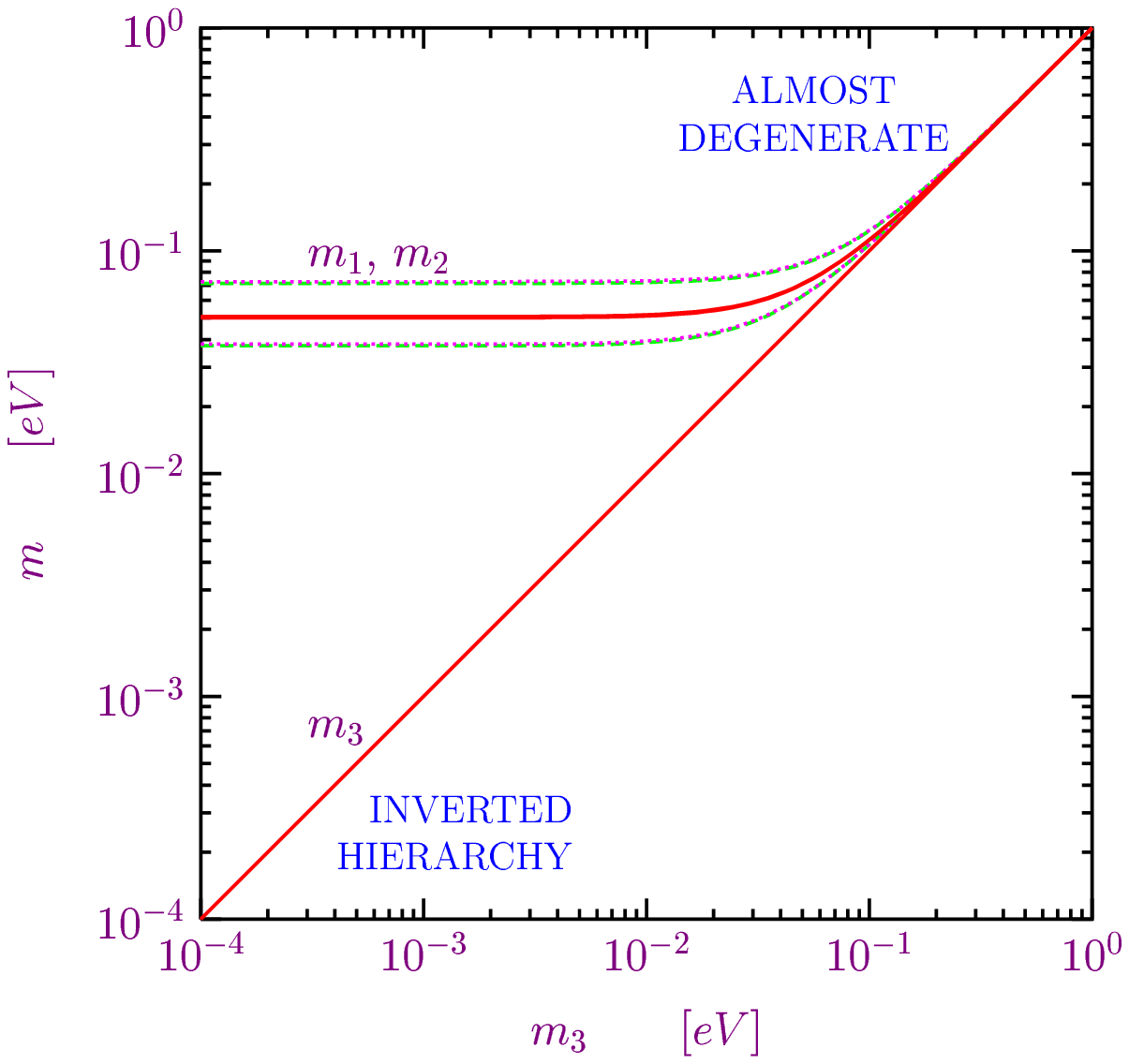}
\end{center}
\end{minipage}
\caption{ \label{3ma}
Allowed ranges for the neutrino masses as functions
of the lightest mass $m_1$ and $m_3$ in the normal and inverted
three-neutrino scheme, respectively.
}
\end{figure*}

In the case of three-neutrino mixing there are no
sterile neutrinos,
in agreement with the absence of any indication in favor
of active--sterile transitions in
both solar and atmospheric neutrino experiments.
Let us however emphasize
that three-neutrino mixing cannot explain the
indications in favor of short-baseline
$\bar\nu_\mu\to\bar\nu_e$
transitions observed in the LSND experiment
\cite{LSND},
which are presently under investigation in the
MiniBooNE experiment
\cite{hep-ex/0210020}.

Let us now discuss the current information on
the three-neutrino mixing matrix $U$.
In solar neutrino experiments
$\nu_\mu$ and $\nu_\tau$ are
indistinguishable,
because the energy is well below $\mu$ and $\tau$ production
and $\nu_\mu$, $\nu_\tau$ can be detected only through flavor-blind
neutral-current interactions.
Hence,
solar neutrino oscillations,
as well as the oscillations in the KamLAND experiment,
depend only on the first row
$U_{e1}$,
$U_{e2}$,
$U_{e3}$
of the mixing matrix,
which regulates $\nu_e$ and $\bar\nu_e$ disappearance.
The hierarchy $\Delta{m}^2_{\mathrm{SUN}} \ll \Delta{m}^2_{\mathrm{ATM}}$
implies that neutrino oscillations generated by
$\Delta{m}^2_{\mathrm{ATM}}$ in Eq.~(\ref{006})
depend only on the last column
$U_{e3}$,
$U_{\mu3}$,
$U_{\tau3}$
of the mixing matrix,
because $m_1$ and $m_2$ are indistinguishable.
The only connection between
solar and atmospheric oscillations
is due to the element
$U_{e3}$.
The negative result of the CHOOZ long-baseline
$\bar\nu_e$ disappearance experiment
\cite{Apollonio:2003gd}
implies that electron neutrinos do not oscillate
at the atmospheric scale,
in agreement with the above mentioned disfavoring
of $\nu_\mu\to\nu_e$
transitions in atmospheric experiments.
The CHOOZ bound on
the effective mixing angle
$
\sin^2 2\vartheta_{\mathrm{CHOOZ}}
=
4 \, |U_{e3}|^2 \left( 1 - |U_{e3}|^2 \right)
$
implies that
$|U_{e3}|$
is small:
$ |U_{e3}|^2 < 5 \times 10^{-2} $
(99.73\% C.L.)
\cite{Fogli:2002pb}.
Therefore,
solar and atmospheric neutrino oscillations are practically decoupled
\cite{Bilenky:1998tw}
and the effective mixing angles in
solar, atmospheric and CHOOZ experiments
can be related to the elements of the three-neutrino mixing matrix by
(see also Ref.~\cite{hep-ph/0212142})
\begin{equation}
\sin^2\vartheta_{\mathrm{SUN}}
=
\frac{|U_{e2}|^2}{1-|U_{e3}|^2}
\qquad
\sin^2\vartheta_{\mathrm{ATM}}
=
|U_{\mu3}|^2
\,,
\label{011}
\end{equation}
and
$
\sin^2\vartheta_{\mathrm{CHOOZ}}
=
|U_{e3}|^2
$
Taking into account all the above experimental constraints,
the best-fit value for the mixing matrix $U$ is
\begin{equation}
U_{\mathrm{bf}}
\simeq
\left( \begin{smallmatrix}
-0.83 & 0.56 &  0.00 \\
 0.40 & 0.59 &  0.71 \\
 0.40 & 0.59 & -0.71
\end{smallmatrix} \right)
\,.
\label{012}
\end{equation}
We have also reconstructed the allowed ranges for the
elements of the mixing matrix
(see Ref.~\cite{hep-ph/0306001}
for a more precise reconstruction
taking into account the correlations
among the mixing parameters):
\begin{equation}
|U|
\simeq
\left( \begin{smallmatrix}
0.71-0.88 & 0.46-0.68 & 0.00-0.22 \\
0.08-0.66 & 0.26-0.79 & 0.55-0.85 \\
0.10-0.66 & 0.28-0.80 & 0.51-0.83
\end{smallmatrix} \right)
\,.
\label{013}
\end{equation}
Such mixing matrix,
with all elements large except $U_{e3}$,
is called ``bilarge''.
It is very different from the quark mixing matrix,
in which mixing is very small.
Such difference is an important
piece of information for our understanding
of the physics beyond the Standard Model,
which presumably involves some sort of quark-lepton unification.

\begin{figure*}
\begin{minipage}[t]{0.47\textwidth}
\begin{center}
\includegraphics*[bb=121 376 465 702, width=0.99\textwidth]{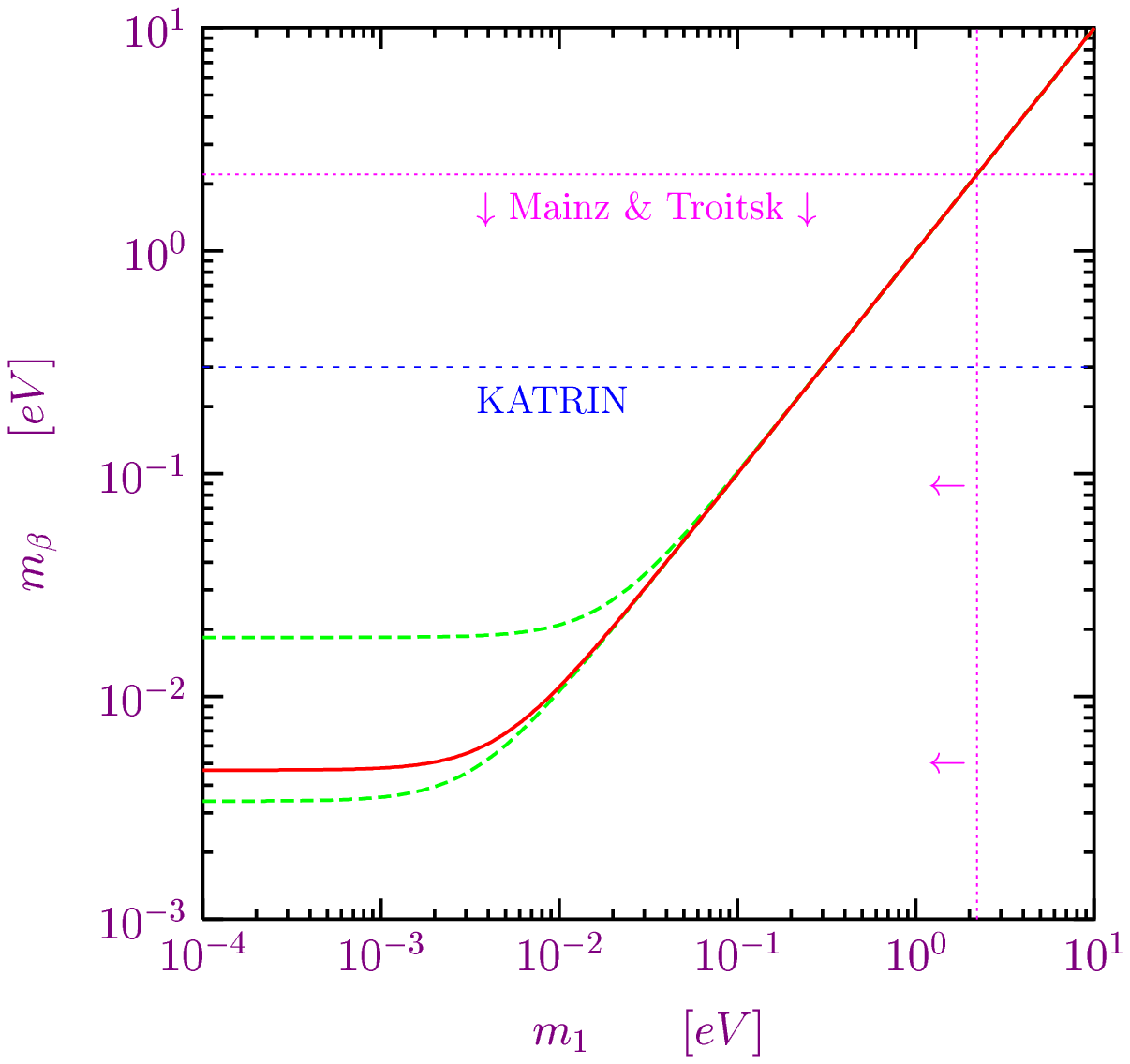}
\end{center}
\end{minipage}
\hfill
\begin{minipage}[t]{0.47\textwidth}
\begin{center}
\includegraphics*[bb=121 376 465 702, width=0.99\textwidth]{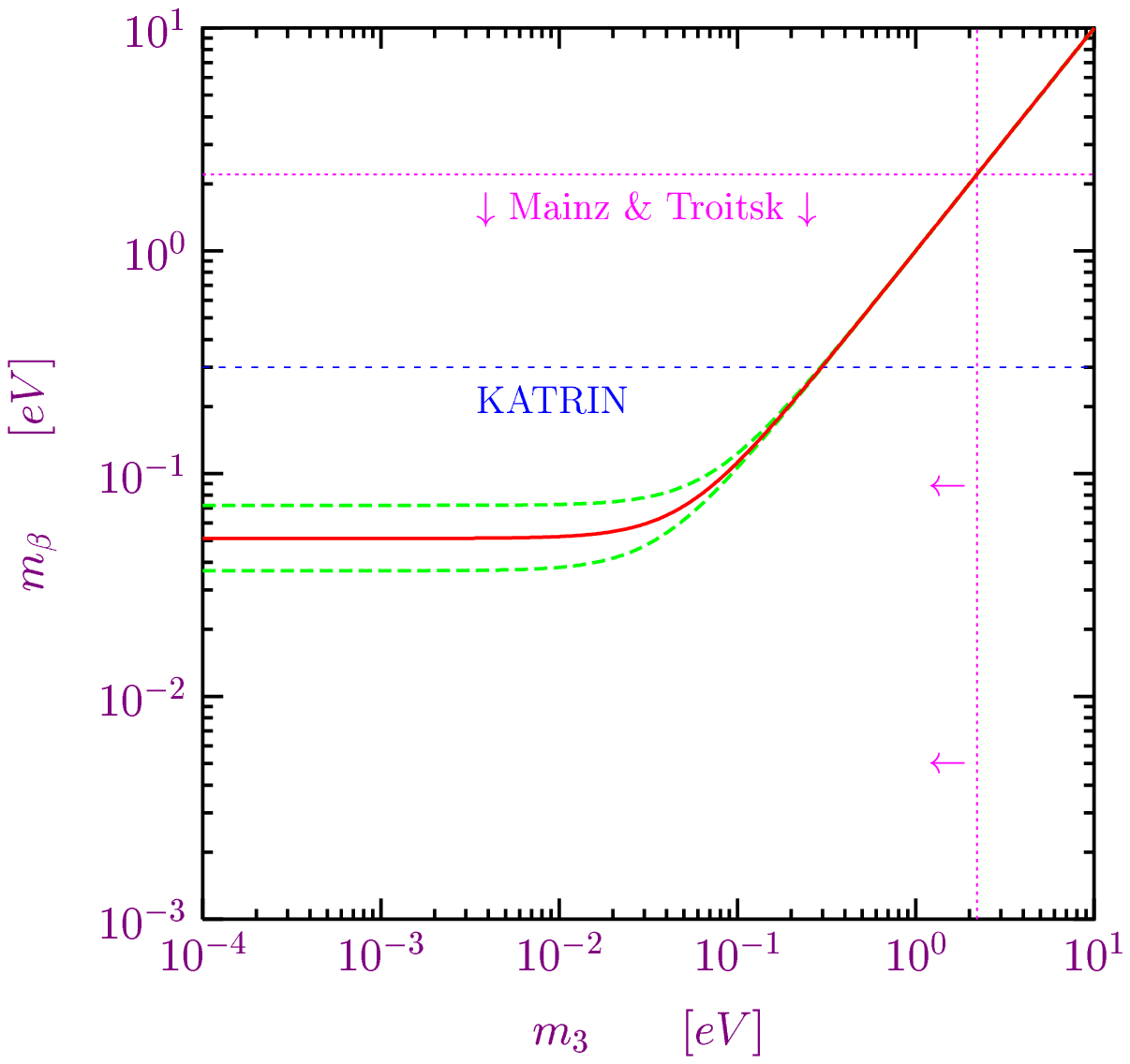}
\end{center}
\end{minipage}
\caption{ \label{mb}
Effective neutrino mass $m_\beta$
in Tritium $\beta$-decay experiments as a function
of the lightest mass $m_1$ and $m_3$ in the normal and inverted
three-neutrino scheme, respectively.
}
\end{figure*}

The absolute scale of neutrino masses
is not determined by the observation of
neutrino oscillations,
which
depend
only on the differences of the squares of neutrino masses.
Figure~\ref{3ma} shows the allowed ranges
(between the dashed and dotted lines)
for the neutrino masses
obtained from the allowed values of the oscillation parameters
in Eqs.~(\ref{001}), (\ref{002}), (\ref{003}), (\ref{004}),
as functions of the lightest mass
in the normal and inverted three-neutrino schemes.
The solid lines correspond to the best fit values of the oscillation parameters.
One can see that at least two neutrinos have masses
larger than about
$7 \times 10^{-3} \, \mathrm{eV}$.

The most sensitive known ways to probe the
absolute values of neutrino masses
are
the observation of the end-point part of
the electron spectrum in Tritium $\beta$-decay,
the observation of large-scale structures
in the early universe
and
the search for neutrinoless double-$\beta$ decay,
if neutrinos are Majorana particles
(see Ref.~\cite{Bilenky:2002aw};
we do not consider here the interesting possibility
to determine neutrino masses through the
observation of supernova neutrinos).


Up to now,
no indication of a neutrino mass has been found in
Tritium $\beta$-decay experiments,
leading to an upper limit on the effective mass
\begin{equation}
m_\beta = \sqrt{ \sum_k |U_{ek}|^2 m_k^2 }
\label{007}
\end{equation}
of $2.2 \, \mathrm{eV}$
at 95\% C.L.
\cite{hep-ex/0210050},
obtained in the Mainz and Troitsk experiments.
After 2007, the KATRIN experiment
\cite{hep-ex/0109033}
will explore $m_\beta$ down to about
$0.2-0.3 \, \mathrm{eV}$.
Figure~\ref{mb} shows the allowed range (between the dashed lines)
for $m_\beta$
obtained from the allowed values of the oscillation parameters
in Eqs.~(\ref{001}), (\ref{002}), (\ref{003}), (\ref{004}),
as a function of the lightest mass
in the normal and inverted three-neutrino schemes.
The solid line corresponds to the best fit values of the oscillation parameters.
One can see that in the normal scheme with a mass hierarchy
$m_\beta$
has a value between about
$3 \times 10^{-3} \, \mathrm{eV}$
and
$2 \times 10^{-2} \, \mathrm{eV}$,
whereas in the inverted scheme
$m_\beta$
is larger than about
$3 \times 10^{-2} \, \mathrm{eV}$.
Therefore,
if in the future it will be possible to constraint
$m_\beta$
to be smaller than about
$3 \times 10^{-2} \, \mathrm{eV}$,
a normal hierarchy of neutrino masses will be established.


The analysis of recent data on cosmic microwave background radiation
and
large scale structure in the universe
in the framework of the standard cosmological model
has allowed to establish an upper bound of
about 1 eV for the sum of neutrino masses,
which implies an upper limit of about 0.3 eV
for the individual masses
\cite{Spergel:2003cb,astro-ph/0303076}.
This limit is already at the same level as the sensitivity of the
future KATRIN experiment.
Let us emphasize,
however,
that the KATRIN experiment is important in order to probe
the neutrino masses in a model-independent way,

\begin{figure*}
\begin{minipage}[t]{0.47\textwidth}
\begin{center}
\includegraphics*[bb=121 376 465 702, width=0.99\textwidth]{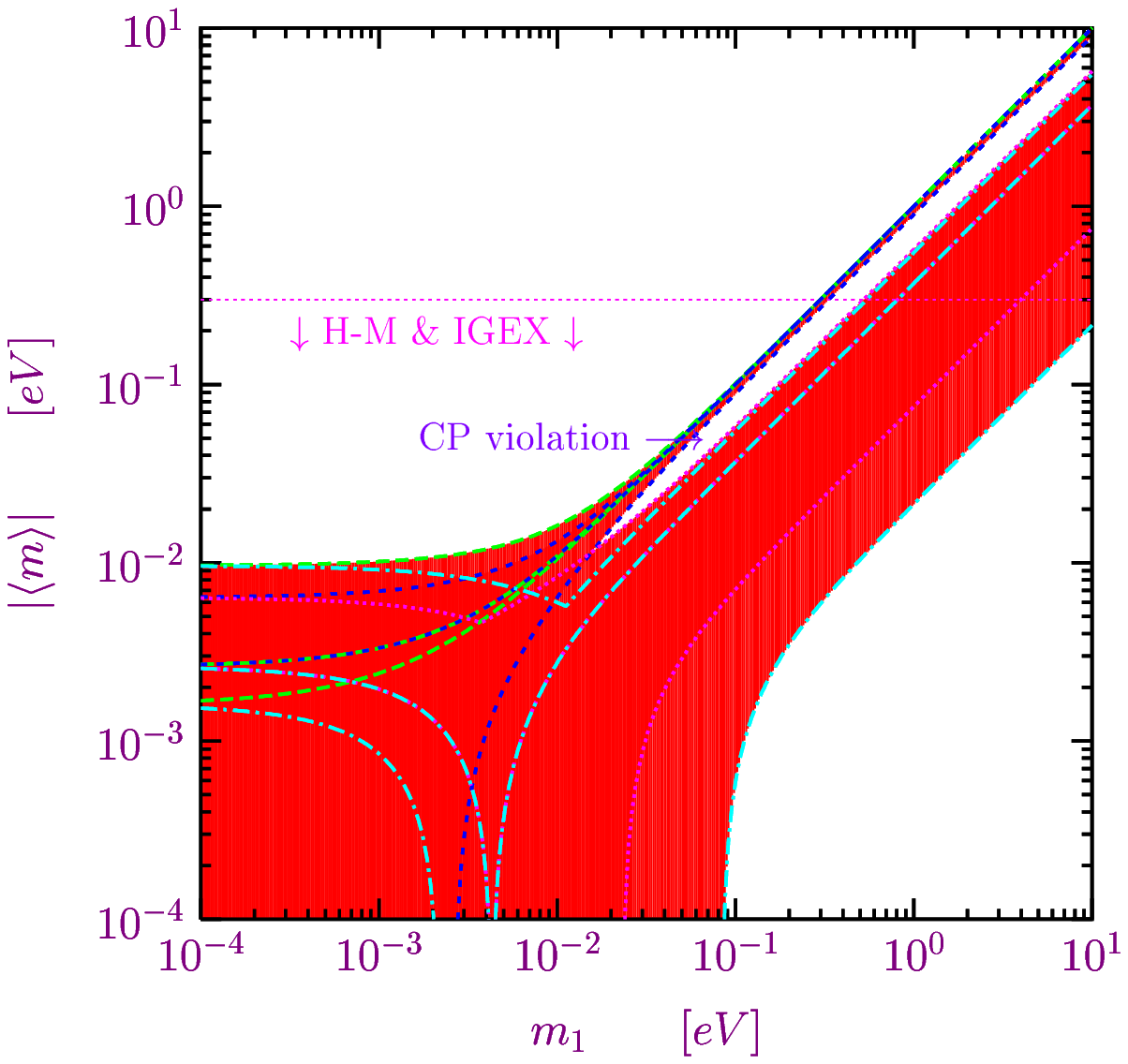}
\end{center}
\end{minipage}
\hfill
\begin{minipage}[t]{0.47\textwidth}
\begin{center}
\includegraphics*[bb=121 376 465 702, width=0.99\textwidth]{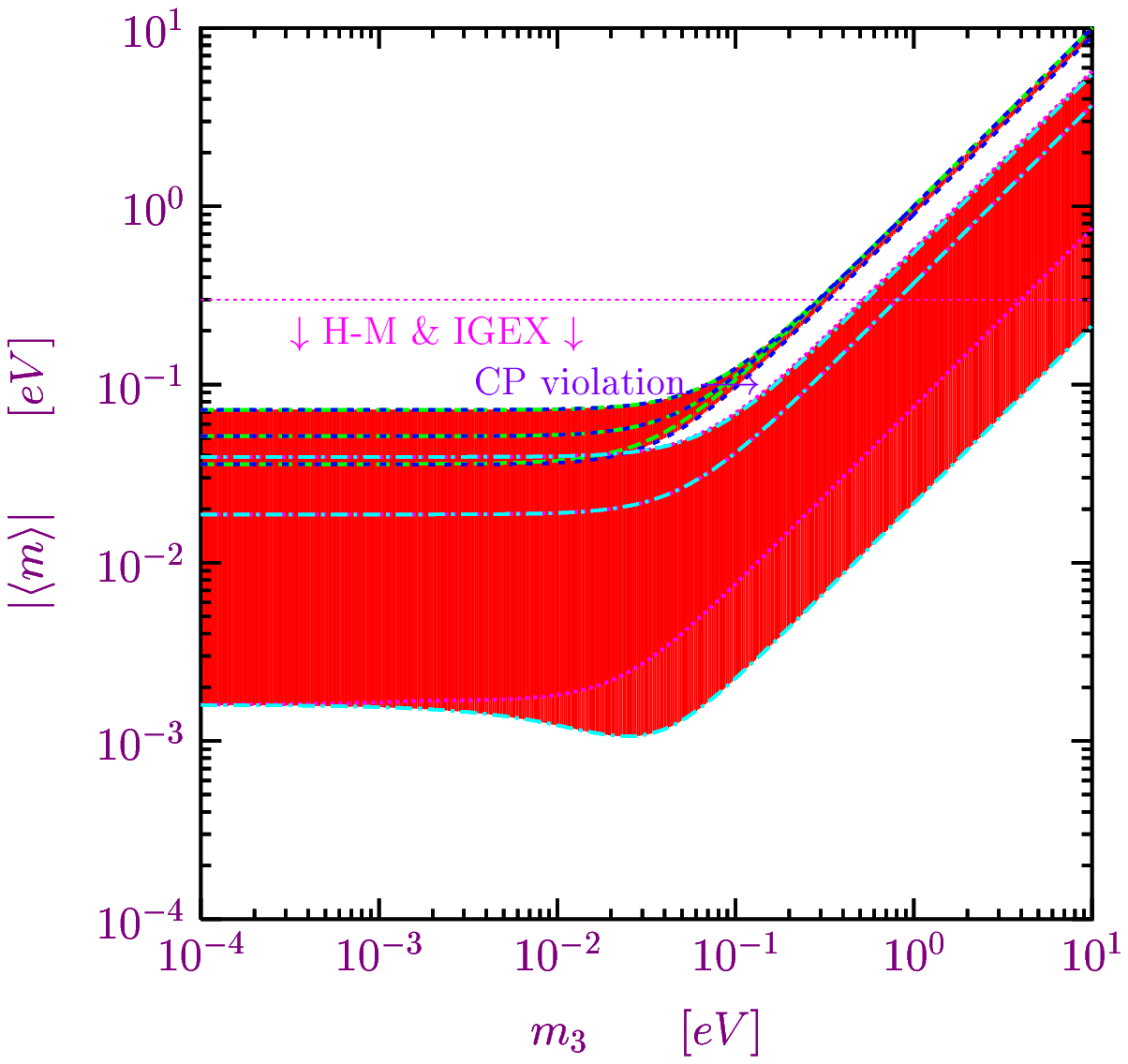}
\end{center}
\end{minipage}
\caption{ \label{db}
Effective Majorana mass $|\langle{m}\rangle|$
in neutrinoless double-$\beta$ decay experiments as a function
of the lightest mass $m_1$ and $m_3$ in the normal and inverted
three-neutrino scheme, respectively.
}
\end{figure*}


A very important open problem in neutrino physics
is the Dirac or Majorana nature of neutrinos.
From the theoretical point of view it
is expected that neutrinos are Majorana particles,
with masses generated by effective Lagrangian terms
in which heavy degrees of freedom have been integrated out
(see Ref.~\cite{Altarelli:2003vk}).
In this case the smallness of neutrino masses
is naturally explained by the suppression due to the
ratio of the electroweak symmetry breaking scale
and
a high energy scale associated with the violation of the total lepton number
and new physics beyond the Standard Model.

The best known way to search for Majorana neutrino masses
is neutrinoless double-$\beta$ decay,
whose amplitude is proportional to the effective Majorana mass
\begin{equation}
|\langle{m}\rangle|
=
\bigg|
\sum_k
U_{ek}^2 \, m_k
\bigg|
\,.
\label{021}
\end{equation}
The present experimental upper limit on $|\langle{m}\rangle|$
between about 0.3 eV and 1.3 eV has been
obtained in the Hei\-del\-berg-Moscow and IGEX experiments.
The large uncertainty is due to the difficulty
of calculating the nuclear matrix element in the decay.
Figure~\ref{db} shows the allowed range for $|\langle{m}\rangle|$
obtained from the allowed values of the oscillation parameters
in Eqs.~(\ref{001}), (\ref{002}), (\ref{003}), (\ref{004}),
as a function of the lightest mass
in the normal and inverted three-neutrino schemes
(see also Ref.~\cite{Pascoli:2002qm}).
If CP is conserved,
$|\langle{m}\rangle|$
is constrained to lie in the shadowed region.
Finding $|\langle{m}\rangle|$ in an unshaded strip
would signal CP violation.
One can see that in the normal scheme large cancellations
between the three mass contributions are possible
and
$|\langle{m}\rangle|$
can be arbitrarily small.
On the other hand,
the cancellations in the inverted scheme are limited,
because $\nu_1$ and $\nu_2$,
with which the electron neutrino has large mixing,
are almost degenerate and much heavier than $\nu_3$.
Since the solar mixing angle is less than maximal,
a complete cancellation between the contributions of $\nu_1$ and $\nu_2$
is excluded,
leading to a lower bound of about
$1 \times 10^{-3} \, \mathrm{eV}$
for $|\langle{m}\rangle|$
in the inverted scheme.
If in the future
$|\langle{m}\rangle|$
will be found to be smaller than
about
$1 \times 10^{-3} \, \mathrm{eV}$,
it will be established that either neutrinos have a mass hierarchy
or they are Dirac particles.
Many neutrinoless double-$\beta$ decay experiments are planned for the future,
but they will unfortunately not be able to probe such small
values of $|\langle{m}\rangle|$,
extending their sensitivity at most in the
$10^{-2} \, \mathrm{eV}$ range
(see Ref.~\cite{Bilenky:2002aw}).


In conclusion,
the recent years have been extraordinarily fruitful
for neutrino physics,
yielding model-independent proofs
of solar and atmospheric neutrino oscillations,
which have provided
important information on the neutrino mixing parameters.
Neglecting the controversial
indications in favor of short-baseline
$\bar\nu_\mu\to\bar\nu_e$
transitions observed in the LSND experiment
\cite{LSND},
three-neutrino mixing
nicely explains all data.
Let us emphasize however that
still several fundamental characteristics of neutrinos
are unknown.
Among them,
the Dirac or Majorana nature of neutrinos,
the absolute scale of neutrino masses,
the distinction between the normal and inverted schemes,
the value of $|U_{e3}|$
and
the existence of CP violation in the lepton sector
are very important for our understanding
of the new physics beyond the Standard Model.



\end{document}